\documentclass[useAMS,usegraphicx,usenatbib]{mn2e}

\usepackage{ulem}
\usepackage{color}
\usepackage{graphicx}
\usepackage{times} 
\usepackage{amssymb}
\usepackage{amsmath}
\usepackage{lscape}
\usepackage{url}
\usepackage{epsfig,bm}
\newif\ifAMStwofonts
\AMStwofontstrue

\def\rg{${\it R}_{\rm g}$}

\def\ka{$K\alpha$}

\def\epicmos1{{\it EPIC}{\rm-MOS1}}
\def\epicmos2{{\it EPIC}{\rm-MOS2}}
\def\epicmos{{\it EPIC}{\rm-MOS}}

\def\bepo{{\it BeppoSAX}}

\def\rxte{{\it RXTE}}

\def\deg{$^{\circ}$}  

\def\ev{\hbox{$\rm\thinspace eV$}}
\def\kev{\hbox{$\rm\thinspace keV$}}

\def\msun{\hbox{$\rm\thinspace M_{\odot}$}}

\def\jb{\hbox{\rm J1655--40}}

\def\vs{\hbox{\rm V4641{\thinspace Sgr}}}

\begin{document}

\title[Misalignment of the microquasar V4641 Sgr (SAX J1819.3--2525) ]
{Misalignment of the microquasar V4641 Sgr (SAX J1819.3--2525) }

\author[R.~G.~Martin, R.~C.~Reis and J.~E.~Pringle] {Rebecca~G.~Martin, Rubens~C.~Reis and J.~E.~Pringle\\
  \footnotesize Institute of Astronomy, Madingley Road, Cambridge CB3
  0HA}

\maketitle

\begin{abstract}
  
  In the microquasar V4641 Sgr the spin of the black hole is thought
  to be misaligned with the binary orbital axis. The accretion disc
  aligns with the black hole spin by the Lense-Thirring effect near to
  the black hole and further out becomes aligned with the binary
  orbital axis.  The inclination of the radio jets and the Fe\ka\ line
  profile have both been used to determine the inclination of the
  inner accretion disc but the measurements are inconsistent.  Using a
  steady state analytical warped disc model for V4641 Sgr we find that
  the inner disc region is flat and aligned with the black hole up to
  about $900\, R_{\rm g}$. Thus if both the radio jet and fluorescent
  emission originates in the same inner region then the measurements
  of the inner disc inclination should be the same.

\end{abstract}


\section{Introduction}

Microquasars are binary systems in which material is accreted from a
normal star on to a compact object. They differ from typical X-ray
binaries by the strong presence of a persistent or episodic radio jet
\citep{MR99}. The compact object is usually associated with a neutron
star or a stellar mass black hole. To date there are over fifteen
microquasars for which the compact object has been dynamically
confirmed to be a stellar mass black hole \citep{O03}. However, only
four of these systems have well resolved relativistic radio jets
\citep[XTE{\thinspace J1550--564}, GRO{\thinspace J1655--40},
GRS{\thinspace 1915+105} and V4641 Sgr,][]{G03}.

It is usually assumed that the inclination of the jet axis is
perpendicular to that of the orbital plane of the binary. However, it
has been shown by \cite{Macc} and more recently by \cite{MTP08} that
the alignment time-scale in microquasars is usually a significant
fraction of the lifetime of the system. If the black hole in such
microquasars were formed with misaligned angular momentum, as expected
from supernova-induced kicks, then it would be likely that the system
would remain misaligned for most of its lifetime.

Precise measurements of both the orbital plane and jet inclination are
known for two systems. GRO{\thinspace \jb} is a microquasar thought to
contain a rapidly rotating black hole \citep{Z97,R08a} with a mass
constrained to be larger than $6.0\msun$ \citep{OB97}. \cite{H95}
measured a jet inclination of $85$\deg$\pm{\thinspace 2}$\deg\ to the
line-of-sight and it is thus misaligned by at least $10$\deg\ to the
orbital plane, with an inclination of $70.2$\deg$\pm{\thinspace
  1.9}$\deg\ \citep{G01}. \cite{MTP08} presented a detailed
investigation of the alignment timescale of this system and found that
it is consistent with the lifetime of the secondary star.

V4641{\thinspace Sgr} was discovered as an X-ray source independently
with the Wide Field Cameras on \bepo\ on 1999 February 20
\citep{intZ99} and with the Proportional Counter Array on the {\it
  Rossi X-ray Timing Explorer} (\rxte) on 1999 February 18
\citep{MSM99}. Spectroscopic observations made between 1999 September
17 and 1999 October 16 led to a mass function $f(M) =
2.74\pm0.12$\msun\ \citep{O01}. From the lack of X-ray eclipses,
combined with the large amplitude of the folded light curve, they
deduced an orbital inclination angle in the range $60$\deg$\le i_{\rm
  orbit} \le 70$\deg$.7$ and mass $8.73 \le M_{\rm BH} \le
11.70$\msun\ for the compact object. Radio observations of \vs\ 
\citep{H00} made during the 1999 September outburst found the jet
expanding at apparent superluminal velocities with a proper motion
ranging between $0.22\arcsec-1\arcsec$ per day. Based on the radio
information, \cite{O01} suggested that that the jet must be highly
beamed and have an inclination along the line-of-sight of $i_{\rm jet}
\le 10$\deg.  This differs significantly from from the inclination of
the binary orbital axis.

X-ray spectral analysis made on \bepo\ observations of \vs\ 
\citep{intZ00} revealed the presence of a strong Fe\ka\ emission with
an equivalent width between 0.3 and 1\kev. They interpreted this as
fluorescent emission using a photo-ionized medium. The presence of
this Fe fluorescent emission was later confirmed by \cite{R02} from a
collection of \rxte\ data.  They found an emission line at about
$6.6$\kev\ with an equivalent width of about $360$\ev.  A more recent
analysis of the source with \rxte\ data obtained during the outburst
of 2003 August 5--17 \citep{MB06} showed the presence of both a strong
Fe\ka\ fluorescent emission line near 6.5\kev\ and a characteristic
Compton hump at about $20$\kev.  If these features are attributed to
the reprocessing of a hard X-ray (powerlaw) continuum by cold matter
in an accretion disc \citep{RN03,RF07} then the degree of broadening
observed implies that the emitting region is very close to the black
hole. The shape of the line profile can then give an indication of
both the the radius of the emitting material from the black hole as
well as the inclination of the inner accretion disc \citep{F89,L91}.
In this way \cite{M02} obtained an estimate for the inclination of the
innermost part of the accretion disc of $43$\deg$\pm{\thinspace
  15}$\deg ($90\%$ confidence).

\section{Misalignment in \vs}

The orbit of \vs\ is thought to be misaligned with the axis of the jet
by about $60^\circ$. The black hole was formed from a supernova which
probably gave a kick to the black hole. Even a small velocity kick can
lead to a large misalignment between the spin axis of the black hole
and the binary orbital axis \citep{BP95}.

The intermediate inclination angle found from the Fe\ka\ line profile
prompted the suggestion of a hierarchy of inclinations \citep{BMP03}
with $i_{\rm orbit}=65$\deg$\pm5$\deg greater than $i_{\rm inner
  disc}= 43$\deg$\pm15$\deg which is greater than $i_{\rm jet} \le
10$\deg.  The \cite{BZ77} mechanism for jet formation uses the
rotational energy of a rapidly spinning black hole. Relativistic jets
are formed in the inner parts of the accretion disc at around
$6\,R_{\rm g}$ \citep{N05} and so must be perpendicular to the inner
disc.  The gravitational radius is defined as
\begin{equation}
R_{\rm g}=\frac{GM_{\rm BH}}{c^2}=1.54 \times 10^6 \left(\frac{M_{\rm BH}}{10.4{\,\rm M_\odot}}\right)\,\rm cm.
\end{equation}
Unless the mechanism for jet formation is very different, we expect
that the jet formation region of the inner disc is at an inclination
of less than $10$\deg.

This is hard to reconcile even with the huge uncertainties in the
measurement of the inclination of the inner disc from the profile of
the Fe\ka\ line \citep{M02}. The inclinations as measured by these two
independent techniques, the Fe\ka\ and radio jets, are found to differ
by at least $43^\circ-15^\circ-10^\circ= 18^\circ$ if we take the
lower limit for the inner disc inclination.

\section{Warped Disc Model}

A warp in an accretion disc around a spinning black hole is driven by
Lense-Thirring precession.  Evidence for this effect, that leads to a
warped accretion disc, can be seen in NGC 4258 where the disc can be
observed by maser emission \citep{M08}. However, in the case of \vs\ 
we cannot observe the disc directly. The inner parts of the disc align
with the black hole by the \cite{BP75} effect and the outer parts are
aligned with the binary orbit because of the angular momentum with
which they accrete as well as tidal effects.  \cite{SF96} found a
steady state solution for the inclination of a disc warped by the
Bardeen-Petterson effect by solving the disc equations of \cite{P92}
for a disc of constant surface density. \cite{MPT07} generalised this
to a power-law density distribution.  We note that it is possible to
have a microquasar which is still misaligned \citep[e.g.  GRO
J1655-40,][]{MTP08} and so we do not need to go into the details of
the stellar evolution here for this system.

The luminosity of the source is
\begin{equation}
L=6\times 10^{-3}\, L_{\rm EDD} = 7.5 \times 10^{35}\left(\frac{M_\odot}{M_{\rm BH}}\right)\,\rm erg\, s^{-1}
\end{equation}
\citep{M02} where $L_{\rm EDD}$ is the Eddington Luminosity and so we
find $L=7.8\times 10^{36}\,\rm erg\, s^{-1}$ with $M_{\rm
  BH}=10.4\,\rm M_\odot$.  The accretion rate is
\begin{equation}
\dot M=\frac{L}{\epsilon c^2}=1.38\times 10^{-9}\,\rm M_\odot \, yr^{-1}
\end{equation}
where we take the accretion efficiency $\epsilon=0.1$.  We use a
steady state disc model with surface density $\Sigma \propto R^{-3/4}$
\citep{SS73}, where $R$ is the radius in the disc. This is the
outermost region (region c) of the Shakura \& Sunyaev disc. The middle
region (b) has $\Sigma \propto R^{-3/5}$ and the steady state shape of
this disc would be almost identical to that of a disc in region c
\citep{MPT07}. The transition radius from region b to c is at $R_{\rm
  bc}=960\, R_{\rm g}$. Note that we have a different definition of
$R_{\rm g}$ to \cite{SS73}.  The inner most region (a) of the Shakura \&
Sunyaev disc has $\Sigma \propto R^{3/2}$ but the transition radius
does not exist here and so the disc in V4641 Sgr has only regions b
and c.  We also take $\nu_1$ and $\nu_2 \propto R^{3/4}$ \citep{WP99}
as used by \cite{MPT07}, where $\nu_1$ is the viscosity corresponding
to the azimuthal shear and $\nu_2$ to the vertical shear.

\cite{MTP08} find the radius up to which the Lense-Thirring effect
dominates the viscous effects in the disc to be
\begin{align}
  R_{\rm warp} & = 1.40 \times 10^9 \,
  \left(\frac{a}{0.7}\right)^{\frac{4}{7}}
  \left(\frac{\alpha_2}{2}\right)^{-\frac{16}{35}} \left(\frac{M_{\rm
        BH}}{10.4\,{\rm M_\odot}}\right)^{\frac{9}{7}} \cr & \times
  \left(\frac{\dot M}{1.38\times 10^{-9}{\,\rm M_\odot \,
        yr^{-1}}}\right)^{-6/35}\,\rm cm,
\end{align}
where $a$ is the dimensionless spin of the black hole and $\alpha_2$
is the dimensionless viscosity parameter associated with the vertical
shear in the disc. For stellar mass black holes there are only a few
sources with determined spins which vary from about $0.4$ to that of a
maximally rotating black hole \citep{MNS07,R08b, Mi08}. We choose
$a=0.7$ in the middle of this range.The ratio of the warp radius to
the gravitational radius of the disc is
\begin{align}
  \frac{R_{\rm warp}}{R_{\rm g}} & = 912 \,
  \left(\frac{a}{0.7}\right)^{\frac{4}{7}}
  \left(\frac{\alpha_2}{2}\right)^{-\frac{16}{35}} \left(\frac{M_{\rm
        BH}}{10.4\,{\rm M_\odot}}\right)^{\frac{2}{7}} \cr & \times
  \left(\frac{\dot M}{1.38 \times 10^{-9}{\,\rm M_\odot \,
        yr^{-1}}}\right)^{-6/35}.
\label{ratio}
\end{align}
We note that this radius is far from the inner edge of the disc as
defined by the innermost stable circular orbit.

The direction of the angular momentum of a disc annulus is given by
$\bm{l}=(l_x,l_y,l_z)$ with $|\bm{l}|=1$. We let $W=l_x+il_y$ where
$i=\sqrt{-1}$ and find the warped disc profile for this model to be
\begin{align}
  W=&\frac{2 \sin (i_{\rm orbit}-i_{\rm jet})}{\Gamma
    \left(2/7 \right)}\frac{(-i)^{\frac{1}{7}}}{
    (7/4)^{2/7}}\left(\frac{R_{\rm warp}}{R}\right)^{1/4}
  \cr &\times
  K_{2/7}\left(\frac{4\sqrt{2}}{7}(1-i)\left(\frac{R}{R_{\rm
          warp}}\right)^{-\frac{7}{8}}\right)
\label{Wwarp}
\end{align}
\citep{MPT07} in the frame of the black hole. The inclination of the
disc at radius $R$ is
\begin{equation}
\theta (R) = \cos^{-1}(l_z)=\cos^{-1}(\sqrt{1-|W(R)|^2}).
\end{equation}
In Fig.~1 we plot the inclination of the disc in the frame of the
black hole in V4641 Sgr.  We see that the disc is flat and aligned
with the black hole out to a radius of about $R_{\rm warp}$, which in
the case of V4641 Sgr is about $900\, R_{\rm g}$
(equation~\ref{ratio}). The region where the Fe\ka\ line is emitted is
likely to be within the innermost $20\,R_{\rm g}$ \citep{F06} so the
inclination measured by the Fe\ka\ method \citep{M02} should be the
same as that of the innermost part of the accretion disc and should
thus be the same as the inclination of the radio jets which are formed
up to around $6\,R_{\rm g}$.

\begin{figure}
  \epsfxsize=8.4cm \epsfbox{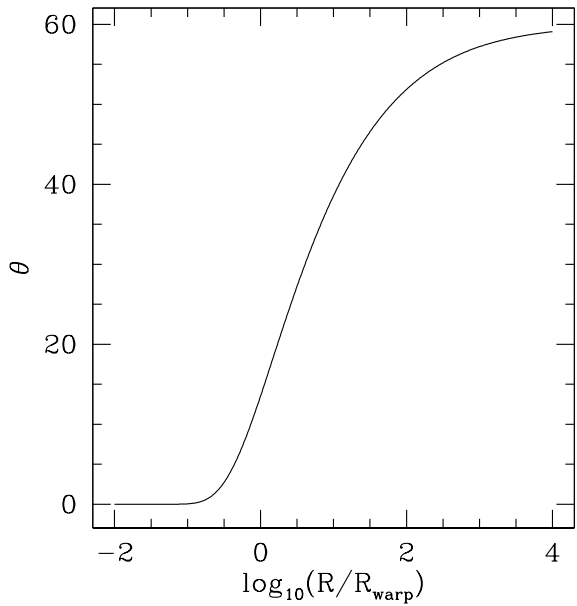}
\caption[]
{ The inclination in the frame of the black hole of a steady state
  accretion disc warped by the Lense-Thirring effect with $\Sigma
  \propto R^{-3/4}$ and $\nu_1$, $\nu_2 \propto R^{3/4}$ which has a
  misalignment of $60^\circ$ between the jet and the binary orbital
  axis}
\end{figure}

\section{Discussion}

In view of the discrepancy between the measured inclination angles of
the innermost parts of the accretion disc by Fe\ka\ profile fitting
and that of the radio jet, \cite{BMP03} suggested a hierarchical
system of inclinations of the jet, inner disc and binary orbit. We
have shown here that if both the jet and iron fluorescent emission
originate from within about $900$\rg\ then their measured inclination
should be the same. This implies that either the measured value from
the jet, Fe\ka\ or both are inaccurate. It was suggested by \cite{C01}
and later by \cite{NC05} that the the inclination of the radio jet
could in fact be as high as that of the orbital inclination. This
uncertainty comes about because the precise time of the radio outburst
is still unknown.  However, the general consensus is that the radio
outbursts started at the same time as those of the X-ray and thus we
get the limit on the inclination of the jet of less than $10$\deg.

The iron emission line seen at $6.5$\kev\ has been generally
interpreted as that originating from a cold accretion disc. In this
way \cite{M02} obtained an inclination of about 43\deg.  However,
\cite{MB06} have argued that the emission could originate in a
varying, optically thick cloud enshrouding the black hole. The strong
broadening of the line would thus be attributed to the highly
dynamical environment and outflows. If this interpretation is correct
then the inclination as measured by \cite{M02} is irrelevant.

We found that the warp radius in \vs\ is around $900$\rg\ for a black
hole with moderate spin and the accretion disc is aligned with the
black hole almost up to this radius. Given the similarity between the
evolutionary state of this system and GRO J1655--40 we expect the spin
of the black hole to be large.  Even if the spin of the black hole is
as low as $a=0.2$ and the accretion rate is as high as the Eddington
accretion rate we find that $R_{\rm warp}=184$ and the disc would
still be essentially flat in this region.  If the jet is emitted
within this radius, as is generally believed, then we would expect an
agreement between the inclination of the jet and that of the inner
disc as obtained from the iron line profile. However it is possible
that the warp radius is very much closer to the black hole in the
unlikely event that the spin of the black hole, $a$, is very small. In
that case, different values for the inclination of the warped inner
disc could be expected.

\section{Conclusions}

We find that, in the accretion disc in V4641 Sgr, the region thought
to be both the origin of the jets and the emission site of the Fe\ka\ 
line is flat and aligned with the central black hole.  Thus we would
expect the inclinations measured for the jets and with the Fe\ka\ line
to be similar.  Because there is a significant difference between the
two measurements we conclude that one or both of them must be
inaccurate or our model incorrect.  It is important that this system
be observed in more detail to resolve this in the near future.

\section*{Acknowledgements}

RGM and RCR thank STFC for financial support.

\end{document}